%% file: jrpelaezMENU07.tex
\documentclass[12pt,a4paper]{conference}

\usepackage{fancyhdr}
\usepackage{graphicx,amsmath,amssymb,cite}
\usepackage{multind}
\makeindex{author} \makeindex{subject}

\input MENU07macros.tex

\begin{document}
%%%%%%%%%%%%%%%%%%%%%%%%%%%%%%%%%%%%%%%%%%%%%%%%%%%%%%%%%%%%%%%%%%%%%%%

\Chapter{Scalar mesons from Unitarized Chiral Perturbation Theory: $N_c$ and quark mass dependences}
           {Scalar meson $N_c$ and $m_q$ dependences}{J.R. Pel\'aez and G. R\'{\i}os}
\vspace{-6 cm}\includegraphics[width=6 cm]{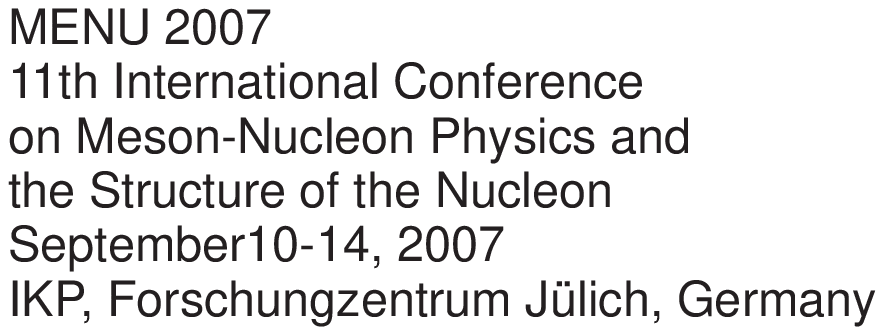}
%\bigskip\bigskip
\vspace{4 cm}

\addcontentsline{toc}{chapter}{{\it Jos\'e R. Pel\'aez}} \label{authorStart}
%%%%%%%%%%%%%%%%%%%%%%%%%%%% NEW SWITCHES %%%%%%%%%%%%%%%%%%%%%%%%%%%%%%

\begin{raggedright}

{\it Jos\'e R. Pel\'aez and G. R\'{\i}os \index{author}{Jos\'e R. Pel\'aez}\\
Departamento de F\'{\i}sica Te\'orica II. Facultad de CC. F\'{\i}sicas.\\
Universidad Complutense\\
28040 Madrid, SPAIN}
\bigskip\bigskip

%%%%%%%%%%%%%%%%%%%%%%%%%%%%%%%%%%%%%%
%%%%%%%%%%%
%%%%%%%%%%%  Repeat for second author
%%%%%%%%%%%
%%%%%%%%%%%%%%%%%%%%%%%%%%%%%%%%%%%%%%
\end{raggedright}

\begin{center}
\textbf{Abstract}
\end{center}
We review recent studies of light scalar meson properties
by means of unitarization techniques, obtained from dispersion theory,
and applied to the Chiral Perturbation 
Theory expansion. In particular, light scalars \index{subject}{light scalar mesons} do not 
follow the $N_c$ dependence of $\bar q q$ states although a subdominant 
$\bar q q$ component may be observed to arise for the $f_0(600)$
 around 1 GeV, where another
$\bar qq$ multiplet is believed to exist. 
Finally, we present
our preliminary results on the quark mass dependence of the $f_0(600)$ and $\rho(770)$ resonances.

\section{Introduction}

Light hadron spectroscopy lies outside the applicability range of QCD
perturbative calculations. Still, in this low energy region one can use 
Chiral Perturbation Theory (ChPT)\cite{chpt1}
\index{subject}{Chiral Perturbation Theory} to obtain a model independent description
of the dynamics of pions, kaons and etas. These particles are the 
Goldstone Bosons (GB) associated to the QCD 
spontaneous breaking of Chiral Symmetry and ChPT is built as 
a low energy expansion that contains those fields in the terms of a Lagrangian
that respect all QCD symmetries, including its 
symmetry breaking pattern. The small quark masses of the three lightest 
flavors can be treated systematically within the perturbative chiral expansion
 and thus ChPT becomes a series in momenta and meson masses, generically $O(p^2/\Lambda^2)$.
 At lowest order there are no free parameters apart from
masses and $f_\pi$, the pion decay constant, that sets the scale $\Lambda\equiv 4 \pi f_\pi$. 
The chiral expansion can be renormalized order by order by absorbing
the loop divergences in higher order counterterms, known as low energy constants (LEC),
whose values depend
on the specific QCD dynamics. That is, other theories with 
spontaneous chiral symmetry breaking at the same 
scale will have the same leading order, but will differ in the values of the 
LEC. 

The renormalized LEC have to be determined from experiment, since
they cannot be calculated from perturbative QCD.  However, thanks to the 
fact that ChPT has the same symmetries than QCD and that it should couple to
different kind of currents in the same way, it is still possible to determine
{\it in a model independent way} how the constants that appear in ChPT,
and therefore the observables, depend on some QCD parameters.
This is indeed the case of the leading dependence on the number of colors $N_c$
and of the dependence on the quark masses.

\section{Unitarization and dispersion theory}

The unitarity \index{subject}{unitarity} of the $S$ matrix implies that, for physical values
of $s$, partial waves $t^{IJ}$
of definite isospin $I$ and angular momenta $J$ for 
{\it elastic} meson-meson scattering should satisfy 
\begin{equation}
{\rm Im }\, t^{IJ}=\sigma \vert t^{IJ}\vert^2 \Rightarrow 
{\rm Im }\,\frac{1}{ t^{IJ}}=-\sigma
\label{unit}
\end{equation}
where $\sigma=2 p/\sqrt{s}$, and $p$ is the CM momenta of the two mesons.
Note that unitarity implies that $\vert t^{IJ}\vert\leq1/\sigma$, and
a strong interaction is characterized precisely 
by the saturation of this unitarity bound. 

However, partial waves are obtained within ChPT 
as a low energy expansion $t\simeq t_2+t_4+t_6+...$,
(To simplify the notation, from now on we will drop the $IJ$ indices.)
where $t_{2k}\equiv O(p/(4\pi f_\pi))^{2k}$,
and thus they cannot satisfy unitarity exactly,
but just perturbatively, i.e:
\begin{equation}
{\rm Im }\, t_2=0, \qquad
{\rm Im }\, t_4=\sigma t_2^2, \qquad {\rm etc...}
\end{equation}

Unitarization \index{subject}{unitarization} methods extend the ChPT series to high energies
by using the fact, remarked in eq.(\ref{unit}),
 that {\it the imaginary part of the inverse amplitude is known exactly}.
Hence, we can approximate the real part of ${\rm Re}\, t^{-1}\simeq t_2^{-2}(t_2+{\rm Re}\, t_4+ ...)$ with ChPT, and find that
\begin{equation}
t\simeq=\frac{1}{{\rm Re}\, t^{-1}-i \sigma}=\frac{t_2}{1- t_4/t_2}
\label{IAM}
\end{equation}
This is known as the one-channel Inverse Amplitude Method \cite{Truong:1988zp,Dobado:1996ps}
\index{subject}{Inverse Amplitude Method}.
A usual complaint is that the 
ChPT series is only valid at low energies, and there is no
reason to use it beyond that regime.

However, not only the complete $t$, but also the {\it inverse amplitude}
 $1/t$ and the ChPT series 
to next to leading order (NLO) and beyond, have an analytic structure
with a ``physical cut'' extending from threshold to $\infty$
and a ``left cut'' from $-\infty$ up to
the highest value of $s$ allowed in the crossed channel ($s=0$ if the two mesons are identical). It is then possible to write the following dispersion 
relation \cite{Dobado:1996ps} \index{subject}{dispersion relation} for $t_4$ 
\begin{eqnarray}
t_4 = b_0+b_1s+b_2s^2+     
\frac{s^3}\pi\int_{s_{th}}^{\infty}\frac{\rm Im\,
t_4(s')ds'}{s'^3(s'-s-i\epsilon)}+LC(t_4).
\label{disp1}
\end{eqnarray}
where ``LC'' stands for a similar integral over the left cut and 
we have three subtractions to ensure convergence. A similar dispersion relation
can be written for the function $G\equiv t_2^2/t$, by simply replacing $t_4$ by $G$
and changing the name of the subtraction constants.
Since $t_2$ is real, these two functions have opposite imaginary parts
on the physical cut so that their contributions from 
the ``physical cut'' integral are {\it exactly} opposite.
Their subtractions constants correspond to the value of these functions
at $s=0$ where it is perfectly justified to use
the ChPT expansion.
And finally, their contributions 
from the left cut are also opposite from each other, 
but this time only up to the NNLO ChPT terms. This guarantees a good description
of the integrand only at low energies, but that is precisely the region that has been weighted
by the three subtractions.

Therefore, note that the IAM is actually exact on the physical region 
and only uses the ChPT approximations for the subtractions constants at $s=0$,
where the use of ChPT is totally justified, and for the left cut, where 
the use of ChPT might not be justified at large $\vert s'\vert$, but
the influence of this region is dumped by the subtractions. The use of the IAM
is even more justified if used sufficiently far from this left cut, since
this has an additional $(s-s')$ suppression. 
In summary, there are no model dependencies in the approach, 
but just approximations to a given order in ChPT, 
and indeed the IAM can be derived not only at NLO
but also at NNLO.

Remarkably, the simple formula of the IAM, eq.(\ref{IAM}), is able to describe 
the $\pi\pi$ and $\pi K$ scattering data not only at low energies,
where it reproduces the ChPT series, but also in the resonance region.
This is done with values of the ChPT parameters that are compatible with 
the values obtained within standard ChPT.

In addition, the IAM generates the poles \cite{Dobado:1996ps,Pelaez:2004xp}
\index{subject}{poles} associated to the resonances in the second Riemann
sheet. This is of relevance since, in particular, the scalar resonances are
the subject of intense debate that has been lasting for several decades, and
as we have seen, the IAM is able to generate their poles from first principles 
like unitarity, analyticity and the QCD chiral symmetry breaking,
 without introducing these resonances by hand. Thus, we can study, without any a priory assumption, the nature 
 of these states as follows from first principles and QCD. 

\section{$N_c$ behavior}

The QCD $1/N_c$ expansion \cite{'tHooft:1973jz} allows for a clear identification of a $\bar qq$ resonance, since
it becomes a bound state, whose width follows an $O(1/N_c)$ behavior, whereas 
its mass should behave as $O(1)$. For our purposes, the relevant observation is that the leading $1/N_c$ behavior
of the ChPT constants is known in a model independent way. Thus, in order to know the leading $N_c$ behavior
of the resonances generated with the IAM, we just have to change the ChPT parameters according to their
established $N_c$ scaling properties. For instance, the pion decay constant scales as $O(\sqrt{N_c})$, and we thus
substitute $f_\pi$ by $f_\pi \sqrt{N_c/3}$. Similar replacements, but according to their respective $N_c$ scaling,
have to be done with all ChPT parameters. 

This procedure \cite{Pelaez:2003dy,Pelaez:2004xp}
was first applied to the coupled channel IAM \cite{Guerrero:1998ei,Pelaez:2004xp}
using the one loop SU(3) ChPT amplitudes, that also include kaons and etas,
and the result was that the light vector resonances $\rho(770)$ and $K^*(982)$ followed remarkably well the expected behavior
of $\bar qq$ states. In contrast, the members of the light scalar nonet, namely, the sigma or $f_0(600)$, the kappa or $K_0(800)$, 
as well as the $f_0(980)$ and $a_0(980)$ resonances, all showed a behavior at odds with that of $\bar qq$ states (Only the $a_0(980)$ could display a $\bar qq$ behavior in a limited corner of parameter space). It follows
that {\it the dominant component of light scalar mesons does not have a $\bar qq$ nature.} 
Other large $N_c$ analysis have reached similar conclusions \cite{otherNc}. 

Very recently\cite{Pelaez:2006nj} we have extended the analysis to two loops in SU(2) ChPT, using the IAM one-channel formalism just described above to NNLO. In addition, we have developed a quantitative measure of how close a resonance is to a $\bar qq$ behavior: Taking into account
subleading orders, to consider a resonance 
a $\bar{q}q$ state, it is enough that 

\begin{equation}
M_{N_c}^{\bar{q}q}=\widetilde{M} \left(1+\frac{\epsilon_M}{N_c}\right),\;
\Gamma_{N_c}^{\bar{q}q}=\frac{\widetilde{\Gamma}}{N_c}
\left(1+\frac{\epsilon_\Gamma}{N_c}\right),
\label{ncqqpoles}
\end{equation}
were $\widetilde{M}$ and $\widetilde{\Gamma}$ are unknown but $N_c$ independent
and the subleading terms have been gathered 
in $\epsilon_M, \epsilon_\Gamma\simeq 1$, since
we expect {\it generically} 30\% uncertainties at $N_c=3$.
Thus, for a $\bar{q}q$ state, 
the {\it expected} $M_{N_c}$ and $\Gamma_{N_c}$ can be obtained
from those at $N_c-1$ generated by the IAM, as follows
\begin{eqnarray}
\hspace*{-.5cm}&&M_{N_c}^{\bar{q}q}\simeq M_{N_c-1}\left[1+\epsilon_M\left(\frac1{N_c}-\frac1{N_c-1}\right)\right]
\equiv M_{N_c-1}+\Delta M^{\bar{q}q}_{N_c},\\
\hspace*{-.5cm}&&\Gamma_{N_c}^{\bar{q}q}
\simeq\frac{\Gamma_{N_c-1}\,(N_c-1)}{N_c}\left[ 1+
\epsilon_\Gamma \left(\frac1{N_c}-\frac1{N_c-1}\right)\right]
\equiv \frac{\Gamma_{N_c-1}\,(N_c-1)}{N_c}+\Delta\Gamma_{N_c}^{\bar{q}q}.
\nonumber
\end{eqnarray}
Note the $\bar{q}q$ index for quantities
obtained {\it assuming} a $\bar{q}q$ behavior.
The reason to write the $N_c$ values from those at $N_c-1$ is to 
be able to  calculate from what $N_c$ a resonance 
starts behaving as a $\bar{q}q$, which is of interest
to search for subdominant $\bar qq$ components.
Thus, we can define an {\it averaged} 
$\bar\chi_{\bar{q}q}^2$%=\chi^{2}_M+\chi^{2}_\Gamma$ 
to check how close a resonance is to 
a $\bar{q}q$ behavior, as follows:
\begin{equation}
  \label{eq:avchi}
  \bar\chi_{\bar{q}q}^2\!=\!\frac1{2n}\sum_{N_c=4}^{n}\!\left[
\left(\frac{M_{N_c}^{\bar{q}q}-M_{N_c}}
{\Delta M_{N_c}^{\bar{q}q}}
\right)^{\!\!2}
\!\!+
\left(\frac{\Gamma_{N_c}^{\bar{q}q}-\Gamma_{N_c}}
{\Delta\Gamma_{N_c}^{\bar{q}q}}\right)^{\!\!2}
\right]
\end{equation}
When this quantity is smaller than one, it indicates a $\bar qq$ behavior, whereas 
a larger value indicates that it does not behave predominantly as such. 
Note also that imposing the minimization of the $\bar \chi^2$ in eq.(\ref{eq:avchi}) we could try to force  a given
$\bar qq$ behavior for a given resonance when fitting data.

When evaluating eq.(\ref{eq:avchi}) above, one has to be careful not to consider too large $N_c$ values.
The reason is that, after all,  we are interested in the physical state at $N_c=3$ and, most likely, 
states are a mixture of different components with different $N_c$ behavior. 
By allowing for too large values of $N_c$ we could be
 altering too radically the nature of the state,
and, since $\bar qq$ states are expected to survive in the large $N_c$ limit, 
whereas other kind of states, like tetraquarks, glueballs, etc... do not, even insignificant
 admixtures of $\bar qq$ at $N_c=3$ could become dominant for sufficiently large $N_c$.
 Thus, the most relevant information will come from $N_c$ not too far from $N_c=3$ and 
 we consider $N_c$ values smaller than one order of magnitude its physical 
value of 3, let us say $N_c<20$.
 
 In this respect, J.A. Oller raised an interesting concern at the end of my talk in this conference,
 about the absence of $\eta'(980)$ in our calculations. Certainly, the $\eta'$ mass is due to the $U_A(1)$ anomaly,
 and decreases like $1/\sqrt{N_c}$. If we were to consider very large $N_c$, this particle should be definitely
 included in our calculations, because it would become the most relevant degree of freedom of QCD at very large $N_c$.
 However, if we limit ourselves to, say $N_c=20$, its mass would be $950 \, {\rm MeV} \times \sqrt{3/20}\simeq 370\, {\rm MeV}$,
 so, it is still much heavier than the pions that can still be considered as the 
 only low energy degrees of freedom of the theory (Furthermore, its contribution would be similar to that of kaon
 loops, although  there 
 are both neutral and charged kaon loops, whereas we only have one $\eta'$).
 This is an additional reason why it would be just wrong to consider too large values of $N_c$ 
 to draw conclusions about the nature of the sigma. 
 
Therefore, with the measure in eq.(\ref{eq:avchi}) and $n=20$,  we are able to 
quantify the deviation of the $f_0(600)$ from the $\bar qq$ behavior: at NLO, even in the most favorable case
when we try to impose that it behaves as a $\bar qq$, the data fit yields a $\bar\chi_{\bar{q}q}^2=125$ for the $f_0(600)$.

When using the NNLO (two loop ChPT calculation), we have many more ChPT parameters, which are not well known,
and give a great deal of freedom. For the NNLO low energy constants (LECS) we thus use 
standard estimates \cite{Bijnens:1997vq} with a $100\%$ uncertainty. Still, there are large correlations 
and some very weak dependences on some of these parameters that can be driven far from their standard values
for negligible improvements in the $\chi^2$ of the data fit. For that reason we stabilize 
the values of the LECS, imposing also the
minimization of a $\chi^2_{LECS}$ together with that of the data. Also, using eq.(\ref{eq:avchi}),
we impose the $\rho(770)$,
which is a well established $\bar qq$ state, to behave as such.
Thus, at two loops, we find that the $f_0(600)$ comes 
out with $\bar\chi_{\bar{q}q}^2=4$ in the most 
favorable case when we try to impose it to behave as a $\bar qq$. Even 
relaxing the $\rho$ $\bar qq$ behavior, we still get $\bar\chi_{\bar{q}q}^2=3.5$ 
for the $f_0(600)$.
In conclusion, {\it the two loop IAM confirms once again that the $f_0(600)$ does not behave predominantly as a
$\bar qq$ state}, whereas that behavior is 
nicely followed by the $\rho$, whose $\bar\chi_{\bar{q}q}^2<0.35$
at NLO and $\bar\chi_{\bar{q}q}^2=0.93$ at NNLO.

\begin{figure}
\begin{center}
\includegraphics[width=8 cm,angle=-90]{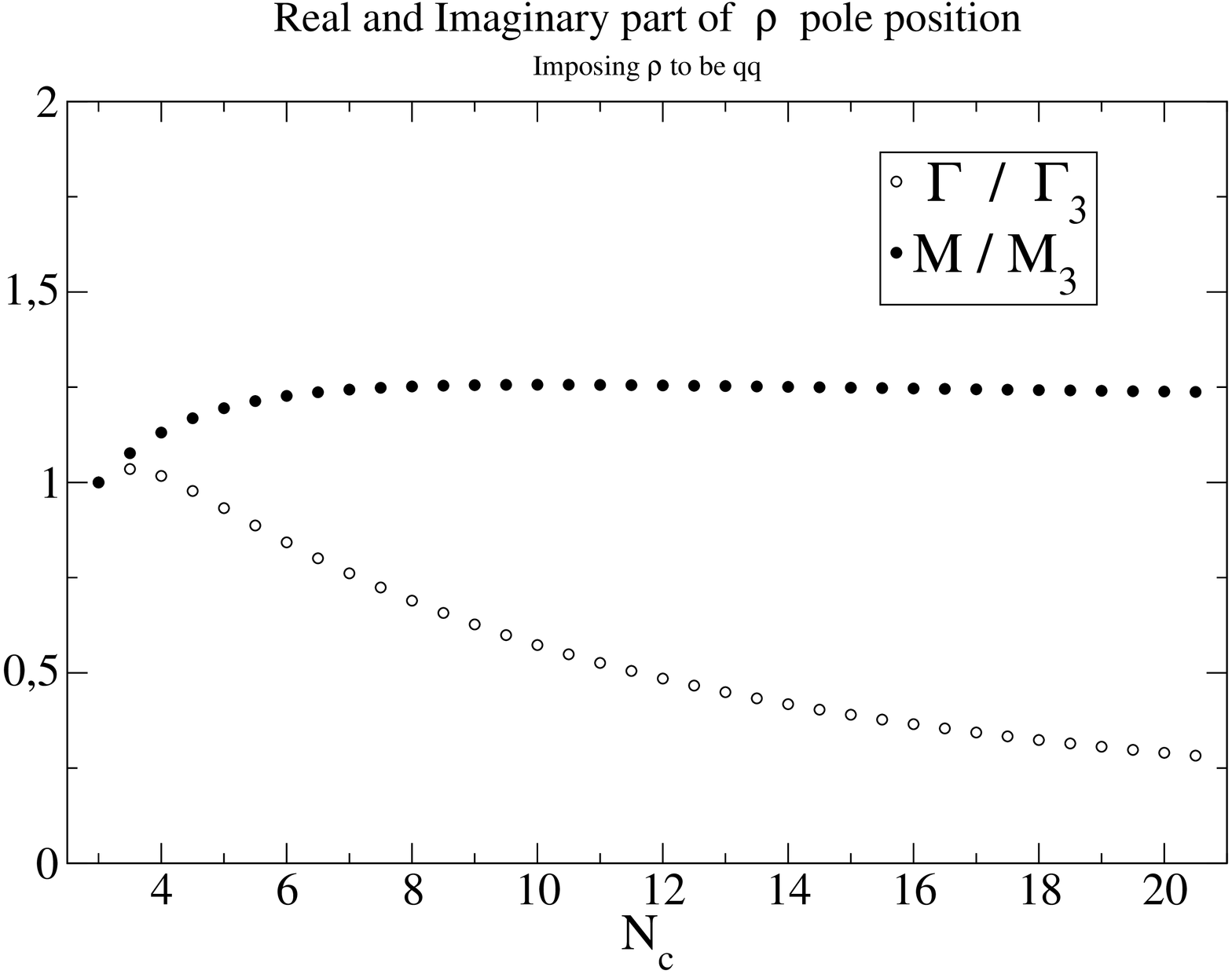}
\includegraphics[width=8 cm,angle=-90]{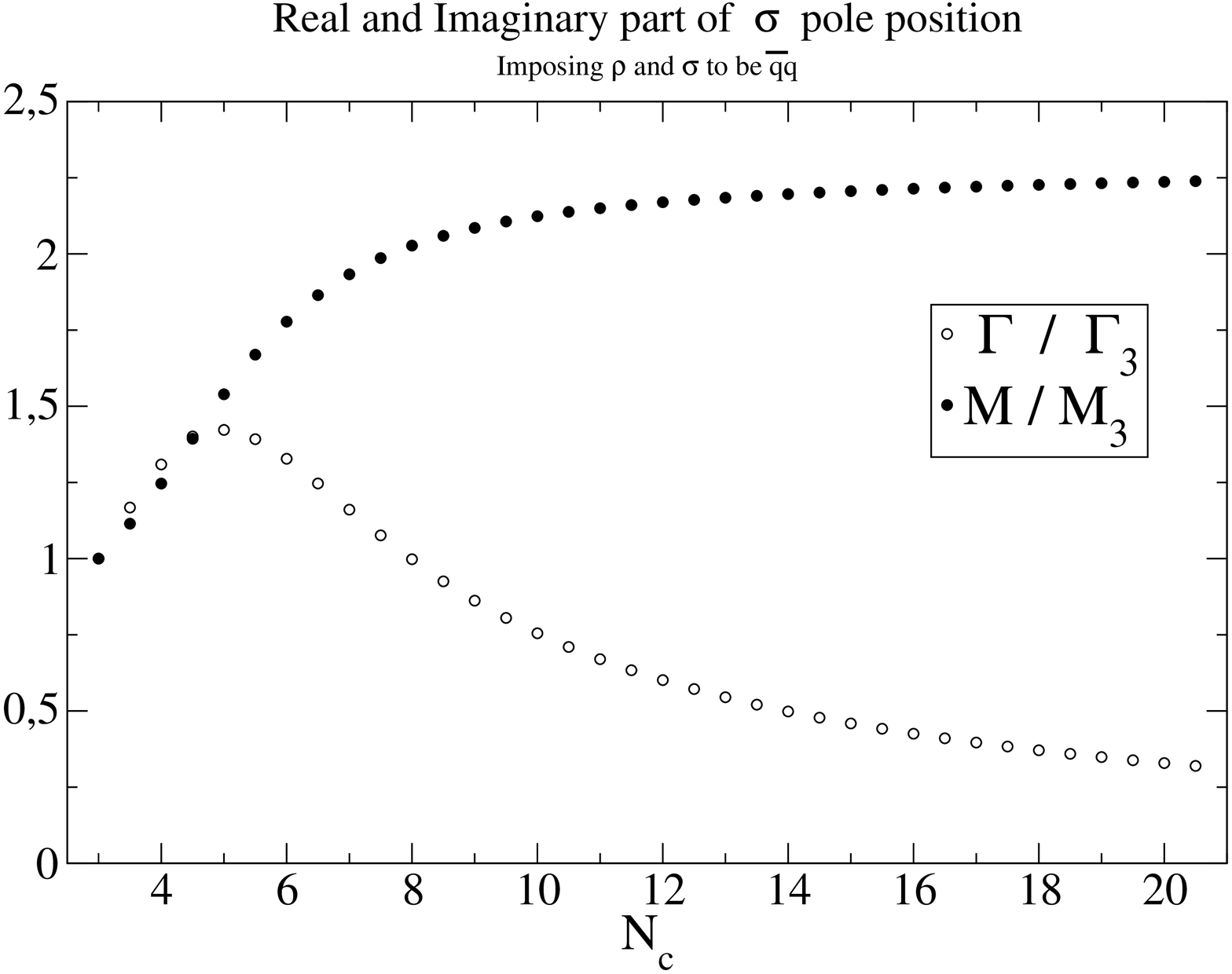}
\caption{Mass and width dependence on $N_c$ for the $\rho$ (Top) and $\sigma$ (Bottom)  resonances. We have normalized all $M$ and $\Gamma$ to their respective physical values at $N_c=3$.
Note that, already for $N_c$ close to 3,
 the $\rho$ mass and width follow nicely the respective $O(1)$ and $O(1/N_c)$
behavior expected for a $\bar qq$ state. In contrast, the $\sigma$ 
does not behave predominantly as such, but only develops, above $N_c=8$ or 10, a 
{\it subdominant} $\bar qq$  behavior with a mass $\simeq 1$ GeV.
 } \label{Fig:F1}
\end{center}
\end{figure}

We show in Fig.\ref{Fig:F1} the NNLO behavior of the $\rho$ and $f_0(600)$ mass and width.
In the top figure, we show the $N_c=3$ behavior of the $\rho$. The full dots represent 
the values, for different $N_c$, of 
its ''pole  mass'', $M$, whereas the empty dots represent its ``pole width'', $\Gamma$, 
both normalized to their
physical values at $N_c=3$. It can be clearly seen that, already for very low $N_c$,
$M/M_3$ starts behaving as $O(1)$
and $\Gamma/\Gamma_3$ as $O(1/N_c)$, as expected for a $\bar qq$ state. 
As commented above, the behavior shown in this plot yields $\bar\chi_{\bar{q}q}^2=0.93$ for the $\rho$.
In contrast, in the bottom figure, and with the same conventions,
  we show the $N_c$ behavior of the $\sigma$ or $f_0(600)$. This time we are also imposing in the
fit minimization that it should behave as a $\bar qq$ state. Obviously it does not,
at least until $N_c\simeq 8$, and its $\bar\chi_{\bar{q}q}^2=4$. Despite the $\sigma$ is still
not behaving predominantly as a $\bar qq$ state, the price due to trying to impose such behavior 
is that the fit suffers a clear deterioration, since
the data now has a $\chi^2/dof=1.5$ (compared to 1.1 before) and the $\rho$ now has
 $\bar\chi_{\bar{q}q}^2=1.3$, thus with a much worse $\bar qq$ behavior.

Remarkably, it is also clear that the $\sigma$ or $f_0(600)$ follows a $\bar qq$ behavior
for $N_c>8$ or 10. This suggests the existence of a {\it subdominant} $\bar qq$ behavior
in the  $\sigma$ or $f_0(600)$ that originates at a mass around twice that of the sigma $\simeq {\rm 1 GeV}$, where a $\bar qq$ nonet is usually located.  This is in good agreement with 
the emerging view that there could be two scalr nonets, a non $\bar qq$ one below 1 GeV
a a $\bar qq$ above \cite{VanBeveren:1986ea}.

\section{Quark mass dependence}

We now present our very preliminary work\cite{inprep} on the quark mass dependence
of the mass and widths of the $\sigma$ and $\rho$ mesons, which could be of interest
to compare with lattice studies, where the small physical masses of light quarks are hard to 
implement.

With the IAM we can also study the quark mass dependence of
the light resonances by changing the meson masses in 
the amplitudes, which is equivalent to change quark masses, since, to leading
order $M_\pi^2\sim m_q+...$.
We have done this in $SU(2)$ Unitarized ChPT at NLO. In 
$\pi\pi$ scattering at NLO there appear four LECs $l_1,\cdots,l_4$.
Phase shifts are almost
insensitive to $l_3^3$ and $l_4^r$, for which we take the values in \cite{chpt1}:
$l_3^r=0.8\pm 3.8,\,l_4^r=6.2\pm 5.7$, evaluated at a
renormalization  scale of $\mu=770\,{\rm MeV}$. 
In contrast, we make a data fit for $l_1^r$ and $l_2^r$, finding $l_1^r=-3.7\pm 0.2,\,l_2^r=5.0\pm 0.4$,
in fairly good agreement with standard values.
Finally, when changing pion masses we have to take into account that amplitudes are
customarily written \cite{chpt1} in terms of the $\mu$ independent LECs $\bar l$ \cite{chpt1}
and the physical pion decay constant 
$f_\pi=f_0\left(1+\frac{M_\pi^2}{16\pi^2 f_0^2}\,\bar l_4+\cdots\right)$
that depend explicitly on the pion mass.

We have taken two criteria to set the applicability limit of our method, that is,
the maximum value of the pion mass we can use. First, we do not want to spoil
the chiral expansion, and second, we do not want the two-pion threshold to reach
the $K \bar K$ threshold. 
Taking into account that  SU(3) ChPT works well
with a kaon
mass of $\sim\,$495 MeV, and that, according to NLO ChPT, 
if we set $M_\pi\simeq 500\,$MeV
 the kaon mass becomes $\sim$ 600 MeV, this means that for 500 MeV pions, $\pi\pi$ scattering is still elastic
 for about 200 MeV above threshold. Hence
the above criteria impose an applicability bound of $M_\pi\simeq$ 500 MeV.
To go beyond that we would need a coupled channel SU(3) formalism.

\begin{figure}
\begin{center}
\includegraphics[width=7.1 cm,angle=-90]{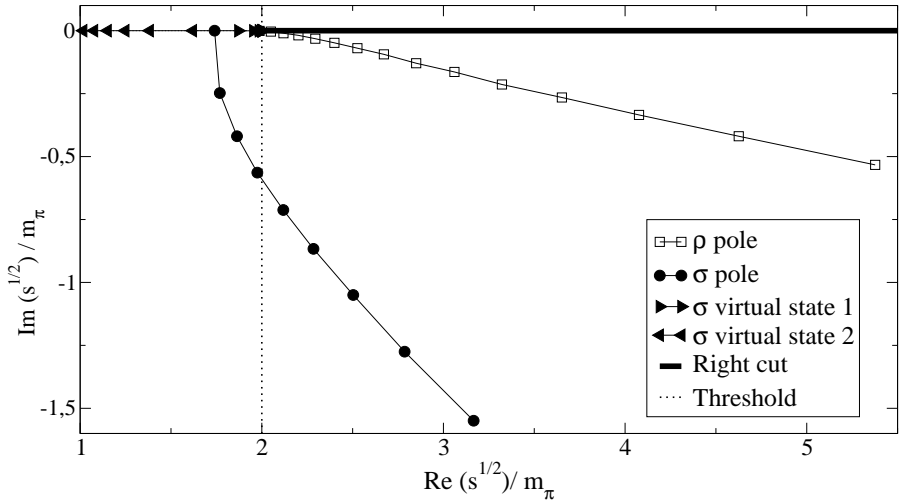}
\caption{$\rho$ and $\sigma$ complex plane pole movement with increasing pion mass.
To ease the comparison of the pole position relative to the two-pion threshold we normalize
 by the pion mass that is changing. Note how the sigma pole moves 
toward the real axis below threshold where it splits in two virtual states,
whereas the $\rho$ pole just moves toward threshold.} \label{Fig:F2}
\end{center}
\end{figure}

We show in Fig.\ref{Fig:F2} how the $\rho $ and $f_0(600)$ poles in the second Riemman sheet
move as $M_\pi$ changes. Both the $\rho$ and $\sigma$
mass increase with the pion mass, but that of the $\sigma$ grows faster.
In addition,  both widths
decrease, partly due to phase space reduction (the two-pion threshold grows
faster than both resonance masses). When the two-pion threshold reaches the
$\sigma$  mass, its pole remains for a short while 
on the second sheet with a non-zero width but quickly 
reaches the real axis where it meets its conjugate partner from the
upper plane and splits again into two poles corresponding to virtual bound states located on the real axis below threshold. 
As the pion mass keeps increasing, one of those "`virtual state"' poles moves toward
threshold and jumps into the first sheet, whereas the other one remains in the second sheet.
Although, of course, this happens for very large $M_\pi$ masses, such an
 analytic structure, with two very asymmetric poles in different sheets of
an angular momentum zero partial wave,
may be a signal of molecular structure, as discussed by M. Pennington in this conference. 

Finally, as the pion mass increases, the $\rho$ pole moves toward the real axis and just when threshold reaches its mass
it jumps into the real axis on the first sheet, thus becoming a traditional bound state,
while its conjugate partner remains
on the second sheet practically at the very same position as the one in the first.

A publication with further details
is in preparation \cite{inprep}
including results of the $f_0(600)$ and $\rho(770)$ mass and width evolution with the pion mass
as well as a comparison with other works and lattice results. Estimates of uncertainties and possibly
an extension to the SU(3) coupled channel case are presently in progress.

% \begin{figure}
% \begin{center}
% \includegraphics[width=8 cm,angle=-90]{mryms.ps}
% \includegraphics[width=8 cm,angle=-90]{wsywr.ps}
% \caption{This is the example of a photograph.} \label{Fig:F2}
% \end{center}
% \end{figure}

%\section{Index}
%Please mark important key words \index{subject}{key word} like
%neutrino mass with \index{subject}{neutrino mass} to have them later
%in an index.

\section*{Acknowledgments}

J.R. Pel\'aez thanks the MENU07 organizers for their kind invitation and the 
great scientific organization. We also thank J.A. Oller and C. Hanhart for useful discussions
and comments.

Research partially funded by Spanish CICYT contracts
FPA2005-02327, BFM2003-00856 as well as Banco Santander/Complutense
contract PR27/05-13955-BSCH, and part of the EU integrated
infrastructure initiative HADRONPHYSICS PROJECT,
under contract RII3-CT-2004-506078. 

\appendix

%\section{References}

%\begin{thebibliography}{000} %for 3 digits
%\begin{thebibliography}{00}  %for 2 digits

 \end{document}

%% file: MENU07macros.tex
\newcommand{\beq}{\begin{equation}}
\newcommand{\eeq}[1]{\label{#1}\end{equation}}
\newcommand{\eeqn}{\end{equation}}

\newcommand{\beqa}{\begin{eqnarray}}
\newcommand{\eeqa}[1]{\label{#1}\end{eqnarray}}
\newcommand{\eeqan}{\end{eqnarray}}

\let\bar=\overbar

\newcommand{\Dslash}{\not{\hbox{\kern-4pt $D$}}}
\newcommand{\dslash}{\not{\hbox{\kern-2pt $\del$}}}

\newcommand{\msb}{{\bar{\ssstyle M \kern -1pt S}}}